\documentstyle[15lomcon,cite,amsmath,amssymb,epsfig]{article}

\def\be{\begin{equation}}
\def\ee{\end{equation}}
\def\bea{\begin{eqnarray}}
\def\eea{\end{eqnarray}}

\def\l{\lambda}

\def\bi{\begin{itemize}}
\def\ei{\end{itemize}}

\begin{document}

\title{VACUUM ENERGY DECAY}

\author{ Enrique \'Alvarez \footnote{e-mail: enrique.alvarez@uam.es},
Roberto Vidal \footnote{e-mail: jroberto.vidal@uam.es}}

\address{Instituto de F\'{\i}sica Te\'orica
UAM/CSIC and Departamento de F\'{\i}sica Te\'orica \\ Universidad
Aut\'onoma de Madrid, E-28049--Madrid, Spain}


\maketitle\abstracts{ The problem of the vacuum energy decay is studied through the analysis of the vacuum survival amplitude ${\mathcal A}(z, z')$. Transition amplitudes are computed for finite time-span, $Z\equiv z^\prime-z$, and their {\em late time}  behavior is discussed up to first order in the coupling constant, $\l$.}

\section{Introduction}
It has been claimed \cite{Polyakov} that the free energy corresponding to an interacting theory in de Sitter space has got an imaginary part that can be  interpreted as some sort of instability. Some general arguments can be advanced supporting that this result is to be found at two-loop or higher orders \cite{Alvarez}. Namely, the optical theorem relates the computation of this imaginary part to a simpler tree level calculation: the vacuum decay into identical particles.\\

It is true, however, that all our intuition is based upon flat space examples, with the ensuing asymptotic regions, and S-matrix elements that can be computed through LSZ techniques. Outside this framework it is not even known how to define a particle to be decayed into, nor the interacting vacuum $|\textrm{vac}\rangle$ in the absence of a well-defined energetic argument. A related issue is the study of the time dependence of transition amplitudes. The linear dependence in time is one of the key aspects of Fermi's golden rule. The fact that is problematic in curved space, where there is no naturally preferred coordinate system in general, has been remarked in \cite{Bros}.\\

We have attempted to calculate some well defined overlaps between states that differ by a finite time transformation in a certain coordinate frame. We have adopted a formalism based on the functional Schr\"odinger picture \cite{Jackiw} applied in curved space-times. We are able to recover some basic results in Minkowski space-time, and also to examine the (conformal) time dependence of transition amplitudes in de Sitter space. For further details we refer the reader to the complete work \cite{Alvarez:2011dc}.

\section{Survival amplitudes}
A \emph{survival amplitude} is an overlap of an state in two different times:
\begin{equation}
{\mathcal A}(t,t_f)=\langle\psi(t_i)|\psi(t_f)\rangle
\end{equation}
The unitary evolution guarantees that the quantity 
\be
\Gamma(t_f,t_i)=-\frac2{t_f-t_i}|{\mathcal A}(t_f,t_i)|
\ee
is positive, and in case it is independent of $T=t_f-t_i$ in the asymptotic regime, could be rightfully interpreted as the \emph{decay width} of the state.\\

We assume that the path-integral representation of the survival amplitude is valid even in a curved background:
\be
{\mathcal A}(t_f,t_i)=\langle \textrm{out}\,t_f|\textrm{in}\,t_i\rangle=\int [D\varphi_f][D\varphi_i]\,\Psi^*_{t_f}[\varphi_f]\Psi_{t_i}[\varphi_i]\,K[J][\varphi_f\,t_f,\,\varphi_i\,t_i]
\ee
where $[D\varphi]$ is the integration measure defined in the space of field configurations at fixed ``time'', the functionals $\Psi$ are the wavefunctionals of the initial and final states, and $K$ is the \emph{field transition amplitude} or Feynman kernel:
\be
K[J][\varphi_f\,t_f\,,\,\varphi_i\,t_i]\equiv\langle\varphi_f\,t_f|\varphi_i\,t_i\rangle\Big|_{J}=\int_{\varphi_i}^{\varphi_f}{\mathcal D}\phi\ e^{i\int_{t_i}^{t_f} L}
\ee
where an external source $J$ is included for computational purposes.\\

This technique can be illustrated with a flat space example. The vacuum survival amplitude for a free field in Minkowski space-time can be rederived directly from the path-integral representation. The functional representation for the vacuum reads:
\be
\langle\varphi|0\rangle=N\,e^{-\frac12\int\varphi\omega\varphi}
\ee 
where $\omega(\vec x,\vec y)$ is the Fourier transform of the energy $\omega_k=\sqrt{k^2+m^2}$. The amplitude can be separated in three pieces\cite{Sakita}: the contribution from the external sources $J$, the integration of the wavefunctionals weighed by the exponential of the classical action, and the determinant of the quantum fluctuations. When $J=0$, the regularization of the two latter contributions yields:
\be
{\mathcal A}(t_f,t_i)=\exp\left[-i V_{n-1} E_0 T+\ldots\right]
\ee
where $V_{n-1}$ is the spatial volume, $E_0$ a constant and we omit subleading contributions independent of time.

\section{Vacuum amplitude in de Sitter space}
\begin{figure}[t]
\begin{center}
\includegraphics[scale=0.5]{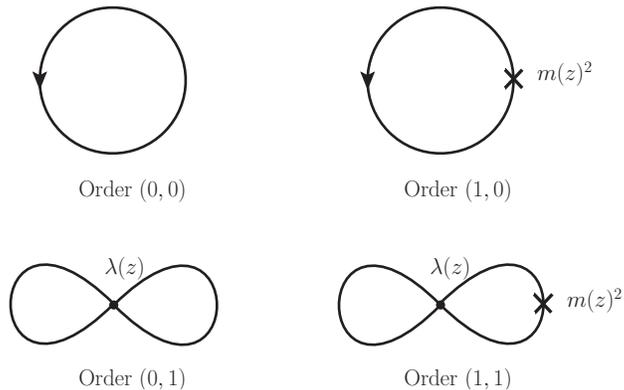}
\caption{The first few diagrams that contribute to the vacuum energy.\label{fig}}
\end{center}
\end{figure}

In a conformal space-time, we can shift all the dependence on the background towards the potential by redefining the field variables:
\be\label{change}
\phi\to a^\frac{n-2}{2}\phi
\ee
where $a$ is the scale factor. In the case of de Sitter in the Poincare patch, $a(z)=1/z$. The survival amplitude can be computed perturbatively in this effective interaction $V(z)$ (made up of time-dependent mass $m(z)$ and interaction coupling $\lambda(z)$) through the diagrams showed in Fig. \ref{fig}. The calculation is similar to the one for Minkowski with a (finite time) propagator modified by the boundary terms coming from (\ref{change}).\\

Putting everything together the (loosely called) vacuum decay width that we obtain exhibits the following behavior:
\begin{align}
&\Gamma (Z)\stackrel{Z\to\infty}{\longrightarrow}\frac{A}{Z}\,\text{Re}I_{00}(Z)+\lambda B Z^{n-5}\,\text{Im} I_{01}(Z)
\end{align}
where $A$ and $B$ are constants and $I_{00}$ and $I_{01}$ are certain integrals that can be computed in the limit $mlz_i\to\infty$:
\begin{align}
&I_{00}(Z)\to-\frac{i \Gamma \left(2-\frac{n}{2}\right) \Gamma \left(\frac{n-1}{2}\right)}{2 \sqrt{\pi }}\nonumber\\
&I_{01}(Z)\to C_n(Z)Z^\frac{1-n}{2}
\end{align}
where $C_n$ is a pure oscillatory function of $Z$. The asymptotic behavior of the decay width in this limit is then proportional to $Z^\frac{n-9}{2}$.

\section*{Acknowledgments}
E.A. thanks Alexander Studenikin for the kind invitation to this interesting and successful conference. This work has been partially supported by the European Commission (HPRN-CT-200-00148) as well as by FPA2009-09017 (DGI del MCyT, Spain) and S2009ESP-1473 (CA Madrid). R.V. is supported by a MEC grant, AP2006-01876.

\end{document}